\documentclass[prl,twocolumn,superscriptaddress]{revtex4}

\usepackage{graphicx}

\usepackage[normalem]{ulem} 
\usepackage{color} 

\begin{document}

\title{Optical coherence of $^{166}$Er:$^{7}$LiYF$_4$ crystal below 1~Kelvin}

\author{N.~Kukharchyk}
\affiliation{Experimentalphysik, Universit\"at des Saarlandes, D-66123 Saarbr\"{u}cken, Germany}

\author{D.~Sholokhov}
\affiliation{Experimentalphysik, Universit\"at des Saarlandes, D-66123 Saarbr\"{u}cken, Germany}

\author{O.~Morozov}
\affiliation{Kazan Federal University, 420008 Kazan, Russian Federation}

\author{S.~L.~Korableva}
\affiliation{Kazan Federal University, 420008 Kazan, Russian Federation}

\author{A.~A.~Kalachev}
\affiliation{Zavoisky Physical-Technical Institute, 420029 Kazan, Russian Federation}

\author{P.~A.~Bushev}
\affiliation{Experimentalphysik, Universit\"at des Saarlandes, D-66123 Saarbr\"{u}cken, Germany}

\date{\today}

\begin{abstract}
We explore optical coherence and spin dynamics of isotopically purified $^{166}$Er:$^{7}$LiYF$_4$ crystal below 1 Kelvin and at weak magnetic fields $<$0.3~T. Crystals were grown in our lab and demonstrate narrow inhomogeneous optical broadening down to 16~MHz. Solid state atomic ensembles with such narrow linewidths are very attractive for implementing of off-resonant Raman quantum memory and for the interfacing of superconducting quantum circuits and telecom C-band optical photons. Both applications require low magnetic field of $\sim10~$mT. However, at conventional experimental temperatures $T>1.5~$K, optical coherence of Er:LYF crystal attains $\simeq10~\mu$s time scale only at strong magnetic fields above 1.5 Tesla. In the present work, we demonstrate that the deep freezing of Er:LYF crystal below 1 Kelvin results in the increase of optical coherence time to $\simeq100~\mu$s at weak fields.

\end{abstract}

\maketitle


\section{Introduction}

Rare-earth (RE) doped solids represent today one of the widely exploited materials for the modern laser and telecommunication industry. Yet, recently, they have raised a strong interest in the field of quantum information storage, signal processing and communication~\cite{Thiel2011}. The RE ions possessing a half integer spin are also known as Kramers ions. The exclusive feature of some Kramers ions, such as Nd$^{3+}$, Yb$^{3+}$ and Er$^{3+}$, is the presence of optical transitions within the standard telecommunication bands, which is very attractive for quantum repeater applications~\cite{GisinRMP2011}. There, RE-doped crystals can be used as quantum memory element for a long-lived storage of entangled photons. 

One of the main challenge associated with using crystals doped with Kramers ions in memory applications is their quite strong unquenched electronic magnetic moment, which in the case of Er$^{3+}$ reaches nearly $8\mu_B$~\cite{Bottger2009}. At weak magnetic fields and conventional temperatures of $T>1.5~$K, large electronic spins mediate a rapid spin-lattice relaxation process which limits the spin coherence time. Another contribution to the decoherence is caused by magnetic dipolar interactions with another electronic spins~\cite{Bottger2006,Lauritzen2008,Baldit2010,Thiel2010,Zambrini2017}. Therefore, in order to attain long coherence time, high magnetic field up to 7~T and low temperatures of 1.5~K are used to polarize an electronic spin bath. By following this prescription the longest optical coherence time of 4.4 ms among solid state systems has been demonstrated for 0.001\% Er$^{3+}$:Y$_2$SiO$_5$ (Er:YSO)~\cite{Bottger2006}. Also, in the very recent experiments M. Ran{\^c}i{\'c} et al. demonstrated optically addressable hyperfine states of $^{167}$Er:YSO with coherence time of 1.3 second~\cite{Sellars2016}. 

In our work, we propose to follow another strategy. The detrimental role of spin-lattice and spin-spin relaxation processes on quantum coherence can be reduced by deep freezing of RE-doped crystals down to ultra-low temperatures, i.e. $T<1~$K. In our previous microwave experiments with erbium doped crystals, it was shown that at sub-Kelvin temperatures substantial polarization of electronic spin bath can be attained already at moderate fields of $\sim0.1~$T. That results in the slowing down of the spin-lattice relaxation rate by orders of magnitude~\cite{Probst2013, Tkalcec2014}. We also found that below 1~K spin coherence time is solely determined by the dynamics of electronic spin bath until it completely freezes out and leaves the only decoherence contribution associated with nuclear spins~\cite{Probst2015}. 

Following up our proposal, we investigate optical coherence of isotopically purified $^{166}$Er$^{3+}$:$^7$LiYF$_4$ (Er:LYF) crystal at sub-Kelvin temperatures and in the magnetic field range of $0-0.3$~T. Since the optical transitions of erbium ions occur between Zeeman sub-levels, the optical coherence is strongly affected by magnetic dipolar interactions with electronic and nuclear spins. Thus, the coherent optical spectroscopy allows us for the exploration of quantum dynamics of erbium spins in a completely new regime, i.e. at sub-Kelvin temperatures and at weak magnetic fields.

Isotopically purified RE-doped LYF crystals are well known for their ultra-narrow optical inhomogeneous broadening $\sim$10 MHz limited by superhyperfine interactions between electronic spins of impurity ions and nuclei spins of the host crystal~\cite{Macfarlane1992,Macfarlane1998}. Today, Er:LYF crystals are again in the focus of the increasing research interest~\cite{Popova2015, Marino2016, Gerasimov2016}, which is driven by their possible applications in broadband quantum memory based on off-resonant Raman protocols~\cite{Walmsley2010,Moiseev:2011cv,Moiseev2013, Zhang2013,Kalachev2013}. The basic idea of such technique is to map the quantum state of incoming optical photon into a long-lived hyperfine spin state, such as coherent spin excitation on a Zero First Order Zeeman (ZEFOZ) transition, which is insensitive to the magnetic field fluctuations~\cite{Longdell2012,Sellars2015}. 

However, the application of ZEFOZ technique for $^{167}$Er doped LYF is quite challenging, because hyperfine ZEFOZ transitions appears only at weak fields of $\sim20~$mT. In the previous works, it was found that Er:LYF crystal requires rather large fields $B>1.5$~T at conventional temperatures of $T\sim1.5$~K for the establishing of optical coherence at $\simeq 10~\mu$s timescale~\cite{Ganem1991,Meltzer1992,Marino2016}. Since the spin coherence of Er:LYF system is determined by dipolar interactions with electronic and nuclear spin baths in the presence of the "frozen core" effect~\cite{Wannemacher1989}, the large field was required for the polarization of erbium spin bath. In the present article, we show that the deep freezing of the Er:LYF crystal below 1 Kelvin leads to an increase of spin coherence time at much weaker fields. Particularly, we show how contribution to decoherence of electronic spins depends on magnetic field and temperature. 

Another motivation of experimenting with narrowband RE-doped crystals at millikelvins is related to the development of a reversible quantum conversion between microwave and telecom C-band optical photons~\cite{Brien2014,Longdell2014,Longdell2015,Blum2015}. Some conversion schemes are based on coherent manipulation of atomic ensembles by using optical pulses. In that case, using of the narrow-line optical ensemble is more favorable since it requires no optical hole burning technique for the efficient transfer/manipulation of the whole spin ensemble~\cite{Afzelius2013}. In addition, the integration of RE-doped solid into a superconducting quantum circuit requires temperatures $T\lesssim 0.1$~K and weak magnetic fields of $\lesssim10~$mT~\cite{Bertet2011}. In that respect, we demonstrate coherent optical spectroscopy of solid-state atomic ensemble at operating temperature range of superconducting quantum circuits.

\section{Experimental setup and sample preparation}

The single crystal $^{166}$Er$^{3+}$:$^7$LiYF$_4$ (0,005 at\%) was grown in our lab by using Bridgman-Stockbarger method in graphite crucibles in argon atmosphere of high purity (99.998\%) at 825 $^\circ$C with a growth rate of 1 mm/h. The source materials $^7$LiF and YF$_3$ with the purity of 99.99\% were produced in Kazan Federal University and were taken with the ratio of 52 mol \% (lithium fluoride) and 48 mol \% (yttrium fluoride). Lithium fluoride was of 99.7 at\% purity of $^7$Li. Erbium-166 isotope was added to the furnace of $^7$LiYF$_4$ as oxide $^{166}$Er$_2$O$_3$ with the isotope concentration of 98\% respectively. The single crystals were grown on a seed crystal with the orientation along the optical axis $a$. After the cleavage and polishing the sample has dimensions of 3$\times$4$\times$5 mm. 

\begin{figure}[ht!]
\includegraphics[width=1\columnwidth]{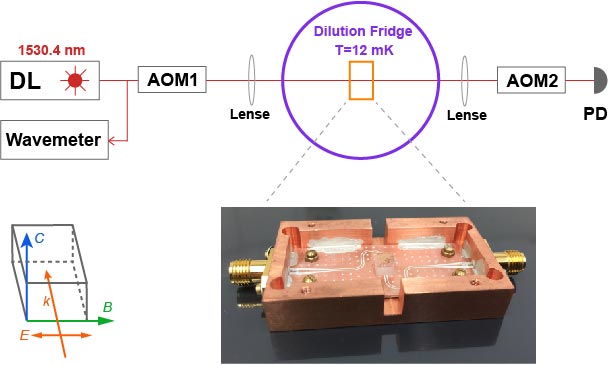}
\caption{(Color online) The experimental setup for studying of optical coherence. DL stands for the diode laser, AOM is the acousto-optical modulator and PD is photoreceiver. Sample mount: the Er:LYF crystal is placed on the top of the microwave substrate. The orientation of the experiment is $B, E, k \perp c $, where $c$ is crystal axis, $E$ is polarization of the beam and $k$ is the wave vector.}
\label{Setup}
\end{figure}

The sketch of experimental setup is shown in Fig.~\ref{Setup}. The crystal is placed on top of the microwave substrate Rogers TMM10i with 50~Ohm transmission line. This line is intended for the future experiments on frequency conversion and long lived spin quantum memory. In order to establish good thermal contact, thin copper plate is located on the top of the crystal and a copper screw presses the crystal to the microwave substrate. In order to preserve good microwave properties of the setup no additional ultra-low temperature thermalization arrangements were made (will be discussed later). The copper sample holder is mounted on the mixing chamber of cryogen-free BlueFors LD-250 dilution refrigerator. The magnetic field up to 0.3~T is created by a pair of superconducting Helmholtz coils and directed perpendicular to the c-axis of the crystal. The field inhomogeneity of our magnet of about 0.3\% is estimated from spin linewidth measurements of Er:YSO crystals. The split-coil magnet is thermally anchored to the still stage of the dilution refrigerator and does not harm the performance of the dilution fridge. Our cryostat has an optical access in the horizontal direction which greatly simplifies transmission spectroscopy. Each radiation shield of the cryostat has two optical windows with gradually reducing diameter starting with 1.5 inch at room temperature down to 0.5 inch at the shield anchored to the cold plate ($\sim$0.1~K). Optical windows are fabricated from Spectrosil 2000 material and AR-coated for the telecom-C wavelength range. The base temperature attainable of the presented cryogenic setup is 12 mK. The full optical setup including laser, pulse preparation and detection setups were installed on the optical breadboard, which was mounted directly on the cryostat's frame. 

\begin{figure}[ht!]
\includegraphics[width=1\columnwidth]{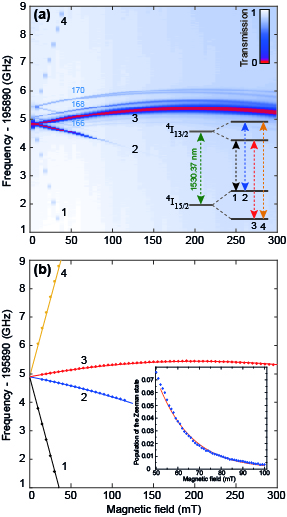}
\caption{(Color online) (a) Optical transmission spectrum of the Er:LYF crystal measured at the base temperature of the cryostat $T=12$~mK as the function of the applied magnetic field. The color bar shows the magnitude of the transmitted signal. The frequency scans are taken in 5~mT steps. The level scheme of Er$^{3+}$ ions in magnetic field shows the observed optical transitions. The weaker absorption signals are attributed to the traces of two additional isotopes $^{168}$Er, $^{170}$Er at $\sim$1 ppm concentration. (b) Dots corresponds to the measured transition frequencies of $^{166}$Er determined from fits of individual absorption lines. Solid lines are the fit by using the Hamiltonian from~\cite{Gerasimov2016}. (Inset) Population of the upper Zeeman state versus the magnetic field. Focused laser beam  warms up erbium spin ensemble to some effective temperature $T_{\textrm{eff}}$. Solid line shows thermal depopulation of Zeeman state at $T_0^{(e)}=90$~mK.}\label{Spectrum2D}
\end{figure}

\section{Transmission optical spectroscopy}

Transmission optical spectroscopy of $^4$I$_{15/2} \leftrightarrow ~^4$I$_{13/2}$ transition at $\lambda=1530.37$~nm (vac) is measured by using a free-running single frequency tunable diode laser Toptica DL-PRO, see Fig.~\ref{Setup}. The laser power is $40~$mW and it has the linewidth of about 0.2~MHz during 1~sec of integration time. The laser frequency is constantly monitored at wavemeter HighFinesse WS6-600 and referenced across P$_9$ transition of fiber coupled $^{12}$C$_2$H$_2$ cell at $\lambda_{P9}=1530.3711$~nm (vac). The laser field is focused with the f=30~cm lens into the crystal, and the beam waist is equal to $\sim 0.1$~mm in diameter. The intensity of the transmitted signal is measured with an amplified InGaAs photoreceiver. 

\begin{figure}[ht!]
\includegraphics[width=1\columnwidth]{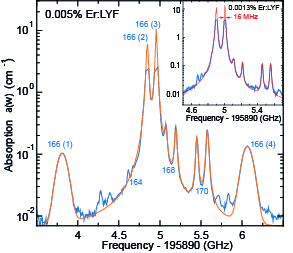}
\caption{(Color online) Absorption coefficient $\alpha(\omega)$ of the Er:LYF crystal extracted from the transmission spectrum at $B=10$~mT. The inset shows the absorption coefficient of 0.0013\% Er:LYF sample.}
\label{SpectrumCut}
\end{figure}

Figure~\ref{Spectrum2D}(a) shows the measured transmission spectrum of 0.005\% Er:LYF crystal as a function of the applied magnetic field. The excitation power of the laser field was set to $0.5~\mu$W. Erbium ions are Kramers ions and they possess a half-integer spin. Every electronic state is split into two Zeeman levels if magnetic field is applied. Due to the high symmetry $S_4$ of the LYF crystal erbium ions substitute yttrium ions only in a single position and no magnetic class splitting (as for Er:YSO) is present. The transmission spectrum consists of 4 lines: transitions 2 and 3 with small g-factor g$_{2,3}=0.4$ and transitions 1 and 4 with larger one of g$_{1,4}=7.7$. Because the Stark levels I$_{13/2}(0)$ and I$_{13/2}(1)$ of the optical excited state are only $\sim 120$~GHz apart~\cite{Marino2016} they mix at higher fields and result in dependence of g-factors on magnetic field. That mixing yields clock transition (CT) for the transition 3 at $B=0.2$~T which is clearly visible on the spectrum. 

The narrow optical inhomogeneous linewidth of our sample allowed us to resolve an isotopic shift. The weaker lines in the spectrum with the same field dependence are attributed to the traces of $^{168}$Er and $^{170}$Er isotopes at $\sim 1$~ppm residual concentration. The transition 3 for $^{170}$Er$^{3+}$ ion is found to split into two lines above 100~mT. This splitting is assumed to arise from the larger distortion of the crystal lattice by this specific isotope. Local distortions may led to the reduction of the local symmetry for erbium ions and appearance of non-equivalent magnetic positions (magnetic classes). 

Transition 2 becomes more faint at higher fields due to thermal depopulation of the upper Zeeman sub-level of the optical ground state. Its population is extracted from the absorption coefficients $\alpha(\omega,B)$ and plot as a function of the applied magnetic field, see the inset in Fig.~\ref{Spectrum2D}(b). The fit of the experimental data to Boltzman distribution yields the actual temperature of erbium spin ensemble $T_0=90$~mK. This measurement shows the preparation of the spin ensemble in a single quantum state at weak magnetic fields, which is crucial for some quantum memory protocols. 

The inhomogeneous broadening $\Gamma$ of optical transitions is also extracted from $\alpha(\omega,B)$ at every recorded field point. Figure~\ref{SpectrumCut} displays the absorption coefficient measured at 10 mT. The absorption spectrum $\alpha(\omega,B)$ consists of many doublets corresponding to optical transitions between Zeeman levels of different isotopes. The orange line shows the fit of the total spectrum to the set of such contributions. Transitions 2, 3 are fit at best with Lorentzians, whereas transitions 1, 4 can be fit well with Gaussians. Full-width-half-maximum (FWHM) linewidth for transitions 2, 3 is $\Gamma_{2,3}/2\pi=35$~MHz and for transitions 1, 4 is $\Gamma_{1,4}/2\pi=130$~MHz. The linewidth of optical transitions is partially determined by super-hyperfine interactions with nuclei spins of the host crystal (mainly with fluorine) and reads as $\Gamma=\Gamma_0+\left(\textrm{g}_e\pm \textrm{g}_g \right) \mu_B \delta B /\hbar+\delta \textrm{g} \mu_B B/\hbar $, where $\Gamma_0$ stands for the intrinsic linewidth due to the crystal strain, g$_{e,g}$ are g-factors of optical ground or excited states, $\delta B$ is the magnitude of the local slow varying magnetic field created by the neighboring nuclei spins, sign $+$ stands for transitions 1 and 4, sign $-$ stands for transitions 2 and 3~\cite{Macfarlane1992,Macfarlane1998}, and the term $\delta\textrm{g} \simeq 0.1$ describes additional inhomogeneity of the field and samples. Evaluation of the full data set for the measured linewidths yields intrinsic $\Gamma_0/2\pi=23$~MHz and $\delta B=1$~mT. We also note here, that FWHM linewidth as narrow as $\Gamma_{2,3}/2\pi=16~$MHz is observed with our sample doped with 0.0013\% Er$^{3+}$ ions, see inset in Fig.~\ref{SpectrumCut}, which is in accordance with the record-narrow linewidth observed by Ch. Thiel, Th. B\"{o}ttger and R Cone.~\cite{Thiel2011}.

\section{Coherent optical spectroscopy}

\begin{figure}[ht!]
\includegraphics[width=1\columnwidth]{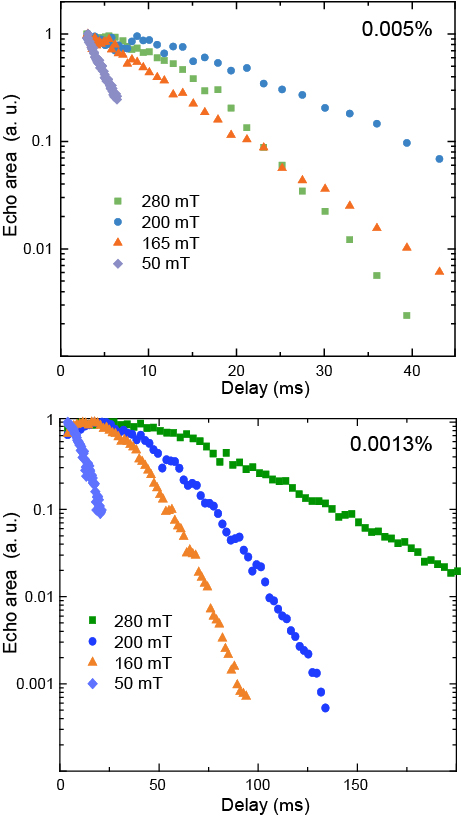}
\caption{(Color online) Normalized echo decay as a function of delay time $\tau$ and magnetic field measured at the base temperature of the dilution refrigerator $T=12~$mK.}
\label{Echo_decay}
\end{figure}

We study spin quantum dynamics of Er:LYF crystal by carrying out two-pulse echo (2PE) experiments. The power level of the laser field at the input of the cryostat is about 10~mW. The typical length of the excitation pulses created by acousto-optical modulator (AOM1) is $\sim$1~$\mu$s. The emitted echo signal ($\sim2$ \% efficiency) is picked up by the AOM2 and detected by the amplified 200 MHz InGaAs photoreceiver. In order to avoid excess heating of the sample, the echo sequence was repeated with relatively slow rate of 1~Hz. At such rate no increase of the temperature of the mixing chamber was detected. The emitted signal was recorded on digitizing oscilloscope and averaged out by using PC. The averaged echo signal was fit at every delay time $\tau$ with $ A \cdot$sech$(t-\tau)$ function and the echo area $A$ was extracted down to the 10$^{-3}$ level. The Fig.~\ref{Echo_decay} represents normalized echo decay measured at the base temperature of the dilution refrigerator at different field values for both 0.0013\% and 0.005\% samples. The displayed signal is normalized to its maximum magnitude. For the entire range of the applied magnetic field the behavior of the echo decay is found to be non-exponential as the consequence of spectral diffusion. Therefore, fit function suggested by W. Mims $I(\tau)=I_0 e^{-2(2\tau/T_M)^x}$ was used in order to extract phase memory time $T_M$ and exponent factor $x$~\cite{Mims1968}.

\section{Magnetic field dependence of optical dephasing}

\begin{figure*}[ht!]
\includegraphics[width=2\columnwidth]{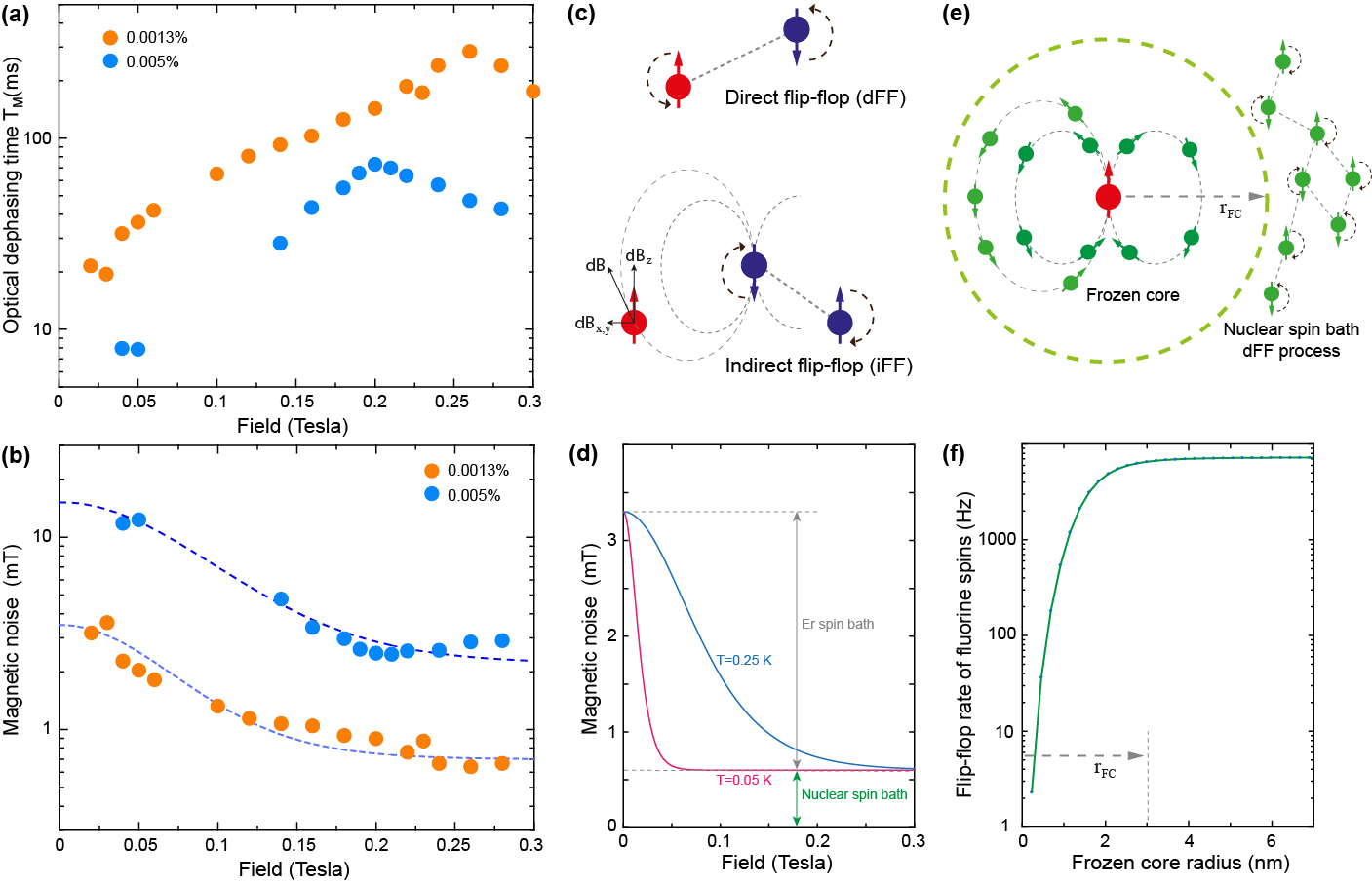}
\caption{(Color online) \textbf{(a)} Optical phase memory time $T_M$ as the function of applied magnetic field measured by using 2PE sequence at the base temperature of the dilution refrigerator $T=12~$mK for 0.0013\% and 0.005\% Er:LYF crystals. Maximal dephasing is attained at turning points of 200~mT and of 250~mT for 0.005\% and 0.0013\% samples respectively. The TP for 0.0013\% sample is shifted due to the crystal missallignment. \textbf{(b)} Magnetic noise $\delta B$ calculated from $T_M(B)$ data by using Eq.~\ref{TM}. Dashed lines represent the fit of data to our model. The fit yields amplitudes of electronic $\delta B_e$ and nuclear $\delta B_n$ magnetic noises as well as the actual temperature of the spin ensemble $T_0$. \textbf{(c)} Contributions of electronic spin bath to optical/spin decoherence due to direct and indirect flip-flops. \textbf{(d)} Magnetic noise calculated by using noise parameters estimated for 0.0013\% sample at temperatures of 0.05~K and 0.25~K. Superhyperfine limit at weak fields can be reached only by the very deep cooling of spin ensemble to tens of millikelvins.\textbf{(e)} Contributions of nuclear spin bath to optical decoherence due to the "frozen core" and flip-flops between nuclear spins in the bulk. \textbf{(f)} Illustration of the "frozen core": calculated flip-flop rate $W^{(n)}(r)$ of fluorine spins by using Eq.~\ref{FF_nuclear} as a function of distance to erbium ion. The region where $W^{(n)}(r)$ drops by 3 orders of magnitude corresponds to the "frozen core".}
\label{Tm_versus_B}
\end{figure*}

The measured dependence of the phase memory time $T_M$ on the applied magnetic field is shown in Fig.~\ref{Tm_versus_B}(a) for both samples. Each data set is measured in a single experimental sequence during 24 hours of the total interrogation time. At zero field and $\tau>3~\mu$s no echo signal is detected. In contrast to the earlier works studying similar systems at high magnetic fields and temperatures above 1.5~K~\cite{Wannemacher1989,Ganem1991, Meltzer1992,Marino2016}, in our experiment, the optical dephasing time attains $\sim10~\mu$s already at weak fields. The echo decay signal experiences strong modulation due to the superhyperfine coupling to nuclear spins~\cite{Mitsunaga1992} at field ranges $70<B<140~$mT for 0.005\% sample and around 70~mT for 0.0013\% sample. Therefore, it is possible to extract dephasing time $T_M$ and exponent $x$ only outside these regions. 

Around 200~mT for 0.005\% crystal and 250~mT for 0.0013\% crystal the magnetic field dependence of $T_M(B)$ shows a typical peak feature associated with reduced sensitivity of optical transition frequency to the local magnetic field fluctuations $\delta B$ in vicinity of the clock transition~\cite{Chaneliere2015}. In order to get insight into dynamical processes of electronic and nuclear spin baths, we determine dependence of $\delta B$ on applied field and temperature from the measured dephasing time. For that purpose, we use a simple model describing behavior of the dephasing time in the vicinity of the CT
\begin{equation}
T_M^{-1}=\vec{S_1}\cdot \vec{\delta B}+  \vec{\delta B} \cdot \hat{S_2} \cdot \vec{\delta B},
\label{TM}
\end{equation}
where $\vec{S_1}$ is the gradient and $\hat{S_2}$ is the curvature of the transition frequency with respect to the magnetic field~\cite{Lovric2011,Longdell2012,Sellars2015,Sellars2016}. In our analysis, we neglect term associated with spin-lattice relaxation, because it is strongly suppressed at our experimental temperature range~\cite{Probst2013, Tkalcec2014}. Our preliminary relaxation time measurements indicate that $T_1^{(SLR)}\sim1-10$~sec. Further, we assume, that the fluctuating field is created by magnetic dipoles which are randomly distributed around the coherently prepared ion, see Fig.~\ref{Tm_versus_B}(c). The total noise magnitude is $\vert \vec{\delta B} \vert = \sqrt{\left(\delta B_x\right)^2+\left(\delta B_y\right)^2+\left(\delta B_z\right)^2}$, where $\delta B_z$ is directed along the applied magnetic field, while $\delta B_y$, $\delta B_x$ are the components of magnetic noise perpendicular to the magnetic field. In addition, $\delta B_x$ is parallel to the c-axis of the crystal. The relation between the components is found by averaging out the dipole-dipole interaction over the uniform angular distribution of magnetic dipoles. Our estimate yields $\delta B_{x,y} \approx 0.5\delta B$ and $\delta B_{z} \approx 0.8\delta B$. 

In the case of Er:LYF crystal, $\vec{S_1}$ is zeroed in $y,z$ directions, i.e. perpendicular to the c-axis. However, along the $x$ direction $S_1^{\parallel}=11.95~$GHz/T. Therefore, it is exactly the magnetic noise component along the $x$ which limits maximum $T_M$ around optical clock transition. 

The substitution of calculated components into Eq.~\ref{TM} yields the magnitude of magnetic noise for each experimental point, which is shown in Fig.\ref{Tm_versus_B}(b). For both samples $\delta B(B)$ demonstrates similar behavior: magnitude of the noise is reduced with increase of the field and stays constant at $B>0.2~$T. In the following, we consider and independently estimate contributions to the noise arising from dipolar coupling between electronic spins of Er$^{3+}$ ions and superhyperfine interaction with nuclear spins of the crystal. 

\subsection{Dephasing due to electronic spins}

There are two main contributions to the decoherence process created by electronic spins, see Fig.\ref{Tm_versus_B}(c). That is namely direct flip-flop process (dFF), which involves the coherently prepared ion and indirect flip-flop (iFF), which involves flips of neighboring spins and thus producing local magnetic field fluctuations $\delta B^{(e)}$ at the position of the coherently prepared ion. 

The electronic spin flip-flop rate in dFF process is estimated by applying the Fermi's "golden rule", s.f. W.B. Mims in ref.~\cite{Geschwind}, as
\begin{equation}
W^{(e)}=\left(\omega_{dd}^{(e)}\right)^2/\Gamma^{(e)},
\label{FFrate}
\end{equation}
where $\omega_{dd}^{(e)}=2.53\mu_0 \textrm{g}_g^2 \mu_B n_e/4\pi \hbar$ is the dipolar linewidth of the electronic erbium spins~\cite{SchweigerJeschke}, $n_e$ is the concentration of erbium spins and $\Gamma^{(e)}=\Gamma_{1,4} \frac{2\textrm{g}_g}{(\textrm{g}_g+\textrm{g}_e)}\simeq 2\pi\times 150~$MHz is the inhomogeneous broadening of an erbium spin ensemble in the ground state. The sample with 0.005\% doping concentration has $n_e\simeq 7 \cdot 10^{17}~$cm$^{-3}$, and estimated electronic spins flip-flop time $T_{ff}^{(e)}=1/W^{(e)}\sim 15~\mu$s. Flip-flop time for more diluted sample with $n_e\simeq 2 \cdot 10^{17}~$cm$^{-3}$ is estimated to be $T_{ff}^{(e)}=1/W^{(e)}\sim 200~\mu$s. 

The above calculated FF times correspond to the case of the fully unpolarized spin bath or high temperature limit. However, the flip-flop rate rapidly decreases with increase of the magnetic field and scales as $W^{(e)}\propto\textrm{sech}^2(\textrm{g} \mu_B B/2k_b T)$, s.f. refs.~\cite{Bottger2006,Takahashi2008,Probst2015}. Therefore, for our diluted samples deeply frozen to temperatures below 1 Kelvin, dFF process seems not to play significant role in the spin dephasing process. The characteristic time scales and inverse linewidths stand in the following order
\begin{equation}
1/\Gamma^{(e)} \ll 1/\omega_{dd}^{(e)} \ll T_M < T_{ff}^{(e)}.
\end{equation}

The magnitude of the local magnetic noise generated in iFF process is contributed by all the flipping spins around the coherently prepared ion and is estimated from $\delta B^{(e)} = \frac{\pi} {9 \sqrt{3}} \mu_0 \textrm{g}_g \mu_B n_e$~\cite{Bupi2009}. The latter yields $\delta B^{(e)} \simeq 12~\mu$T for 0.005\% sample and $\delta B^{(e)} \simeq 3~\mu$T for 0.0013\% sample. Such noise level is created by fully unpolarized spin bath, and as in the case with dFF process, the iFF magnetic noise temperature and field dependence is described by, see~\cite{Bottger2006,Takahashi2008,Probst2015}
\begin{equation}
\delta B^{(e)}(T,B)=\delta B^{(e)}\textrm{sech}^2(\textrm{g} \mu_B B/2k_bT).
\label{dB_TB}	
\end{equation}
At sufficiently large fields or sufficiently low temperatures the electronic spin bath experiences significant polarization and thus the magnitude of iFF magnetic noise $\delta B^{(e)}(T,B)$ will drop to zero. Contrary to that, our measured data shows a constant offset, below which the noise level does not drop. This offset is associated with the superhyperfine limit, i.e. with the local fluctuating field created by the nuclear spins of the host crystal itself. At our experimental conditions the Zeeman levels of nuclear spins are nearly equally populated, hence, the magnetic noise $\delta B^{(n)}$ arising from nuclear spin flip-flops is assumed to stay constant over the experimental field range, see also Fig.~\ref{Tm_versus_B}(d). 

The dashed lines in Fig.~\ref{Tm_versus_B}(b) represent the fit with a function, which takes into account electronic and nuclear spin noise contributions 
\begin{equation}
\delta B = \delta B^{(e)}(T,B)+\delta B^{(n)}.
\label{dB_sum_TB_n}	
\end{equation} 
The fit yields magnitudes of electronic spin noise $\delta B^{(e)}=13,3~\mu$T, magnitudes of nuclear spin noise $\delta B^{(n)}=2,0.7~\mu$T and the actual temperatures of the spin ensembles $T_0=0.2,0.25~$K for 0.005\% and 0.0013\% samples respectively. The contribution of electronic spin bath to spin dephasing at weak fields is found to be dominant even for the diluted (0.0013\%) sample. The electronic spin noise $\delta B^{(e)}$ scales linear with concentration and is in a good agreement with values which is estimated above.

The actual temperatures of spin ensembles $T_0$ in pulsed measurements exceeds the $T_0$ detected in cw measurements. This fact can be explained by a larger heat flow generated by the strong laser pulses when it propagates through the crystal. In order to illustrate the influence of $T_0$ on magnetic noise, we simulate two curves by using Eq.~\ref{dB_sum_TB_n} and taking the noise parameters for 0.0013\% doped Er:LYF crystal at temperatures of 0.05~K and 0.25~K. The curves are displayed in Fig.~\ref{Tm_versus_B}(d). It is clearly seen, that at $T_0<50~$mK the superhyperfine limit can be attained already at the field of 50~mT. 

\subsection{Dephasing due to nuclear spins}

Magnetic noise of a nuclear spins arises from the flip-flops inside the "frozen core" and from the spins in the bulk, see Fig.~\ref{Tm_versus_B}(e). The large magnetic moment $\sim 4 \mu_B$ of Er$^{3+}$ ion creates spatially inhomogeneous field which detunes the resonant frequency of neighboring nuclear spins with respect to each other. This results in strong dependence of nuclear flip-flop rate $W^{(n)}$ on the distance to the optically active erbium ion. The part of nuclear spins in the direct proximity to the electronic spin will form so called "frozen core" with inhibited nuclear spin flip-flop rate. By using similar approach as for electronic spins, we calculate the nuclear flip-flop rate at the distance $r$ from erbium ion as
\begin{equation}
W^{(n)}(r)=\left(\omega_{dd}^{(n)}\right)^2/\Gamma^{(n)}(r),
\label{FF_nuclear}
\end{equation}
where $\omega_{dd}^{(n)}$ is the dipolar linewdith of the nuclear spins, and $\Gamma^{(n)}(r)$ is the inhomogeneous broadening induced by the field gradient $\nabla B$ such that $\Gamma^{(n)}(r)=\Gamma^{(n)}_{bulk}+\gamma_n \nabla B a$, where $\Gamma^{(n)}_{bulk}/2\pi\approx20~$kHz is the linewidth of nuclear spins in the bulk measured in conventional NMR experiment at 2 Kelvin, $\gamma_n$ is the gyromagnetic ratio of nuclear spins and $a$ is the mean distance between nuclear spins.

Fig.~\ref{Tm_versus_B}(f) shows calculated flip-flop rate $W^{(n)}(r)$ of fluorine ions as a function of erbium-fluorine distance. For the simplicity we assume that erbium ion creates a spherically symmetrical field $B(r)=2.18\mu_0 \textrm{g}_g  \mu_B/4\pi r^3$~\cite{Bupi2009}. The mean distance between fluorine ions is equal to $a\approx0.26~$nm. As it is seen from the considered figure, the flip-flop rate increases nearly by 4 orders of magnitude at the distance of several nanometers and reaches a constant value of 7~kHz above the "frozen core" radius of $r_{FC} \sim 3~$nm. There are about 4000 fluorine ions inside the sphere of radius $r_{FC}$ with a strongly suppressed flip-flop dynamics. The radius of the "frozen" was previously estimated by W. B. Mims, see ref.~\cite{Geschwind}. His estimate yields the same $r_{FC}=\sqrt[3]{\gamma_e/(2 \gamma_n n_n)}\sim 3~$nm, where $\gamma_{e,n}$ are the electronic and nuclei gyro-magnetic ratios and $n_n$ is the concentration of nuclear spins.

The local magnetic field created by the nuclear spins sitting at the first coordination sphere is estimated to be around $B_{FC}^{(1)}\simeq 0.7~$mT, see ref.~\cite{Macfarlane1998}. The magnitude of field fluctuation within the time interval of $\tau\sim 100~\mu$s is estimated to be $\delta B_{FC}^{(1)}\sim B_{FC}^{(1)} W^{(n)}(r) \tau \simeq 0.1~\mu$T. The contributions of fluorine ions at the 2-nd and 3-rd coordination spheres are slightly larger and are estimated to be around $0.2~\mu$T. More distant coordination spheres create progressively lower field which scales as $\propto r^{-3}$ with a distance to every erbium ion. In total, the fluorine "frozen core" is estimated to contribute of $\delta B_{FC}\sim1~\mu$T within time interval of $100~\mu$s. Our crude estimation of nuclear spin noise is in reasonable agreement with the observed superhyperfine limit for both samples. The noise contributions of "frozen cores" of Li and Y are significantly lower and therefore are neglected in our consideration. 

Another contribution to the superhyperfine magnetic noise arises from nuclear spins in the bulk, i.e. from those which are outside the frozen core. In this case, the bulk nuclear spins relaxes via magnetic dipole-dipole interaction with their neighbors. Nuclear spin-lattice relaxation path can be neglected because at 2~K it is measured to be $T_1^{(n-SLR)}\sim0.5~$hours. For the case of a cubic lattice, which is good approximation for the high symmetry LYF crystal, the contribution of the bulk nuclear spins to the phase memory of electronic spins $T_M^{s}$ was calculated by I.~M.~Brown, see ref.~\cite{Kevan}, p.202. Since $1/T_M^s \simeq \gamma_e \delta B_{bulk}$, for the fluctuating field we get
\begin{equation}
\delta B_{bulk}=\frac{0.37 \mu_0}{4\pi}\gamma_n \hbar \sqrt{\frac{\gamma_n}{\gamma_e}}\left[I(I+1)\right]^{1/4}n_n.
\label{dB_bulk}
\end{equation}  
Our estimate yields $B_{bulk}^F\simeq 0.9~\mu$T, $B_{bulk}^{Li}\simeq 0.12~\mu$T and $B_{bulk}^Y \simeq 20~n$T for the fluorine, lithium and yttrium spin baths respectively. Although the estimated bulk noise is close to the measured hyperfine noise level, we nevertheless think that the "frozen core" plays more dominant role in the dephasing process. Such conclusion can be derived by studying the behavior of the exponent factor $x$. 

The factor $x$ characterizes the shape of an echo decay and its increase above 1 was interpreted as slowing down of nuclear flip-flop rate during the formation of the "frozen core" around coherently prepared ion~\cite{Ganem1991,Meltzer1992,Szabo1994}. In theoretical work~\cite{Hu1974}, the increase of $x$ is explained by the slowing of flip-flop rare $W$ of B-spins such that the condition $W\tau\ll1$ becomes fulfilled. It was also found that echo decay shows universal characteristic above the certain field regardless of type of the crystal system and doping ions~\cite{Ganem1991}. 

In our experiments with dense ensembles (0.005\% doping concentration), exponent factor $x$ grows up from 0.8 at 50~mT reaching 2 at 280~mT. At weak fields of $<0.1~$T, the electronic spin noise dominates. The condition $W^{(e)}\tau\sim 1$ results in $x\lesssim1$. At fields above $0.1~$T, magnetic field freezes the dynamics of erbium spin bath, and dephasing process is completely determined by the nuclear spins in the "frozen core". In this field range $W^{(n)}\tau\ll1$ and the exponent factor exceeds unity and grows up towards higher fields. For the diluted sample the situation is quite different. Here, for both electronic and nuclear spin baths the condition $W\tau\ll1$ is fulfilled for the entire range of the applied magnetic field therefore exponent factor $x$ is always above 1.5. Thus, our findings also relate the appearance of non-exponential echo decay to the formation of the "frozen core", see also~\cite{Sellars2015,Sellars2016}.
 
The dependence of nuclear noise $\delta B^{(n)}$ on erbium concentration can qualitatively be explained by using "frozen core" concept. Larger concentration of electronic spins creates larger regions where their dipolar fields are cancelled. In this regions nuclei spins flips much faster thus producing larger magnetic noise. We also note here, that influence of electronic spin concentration on coherence time of nuclear spins inside the "frozen core" was observed for ruby~\cite{Szabo1990}.

\section{Temperature dependence of optical dephasing}

\begin{figure*}[ht!]
\includegraphics[width=2\columnwidth]{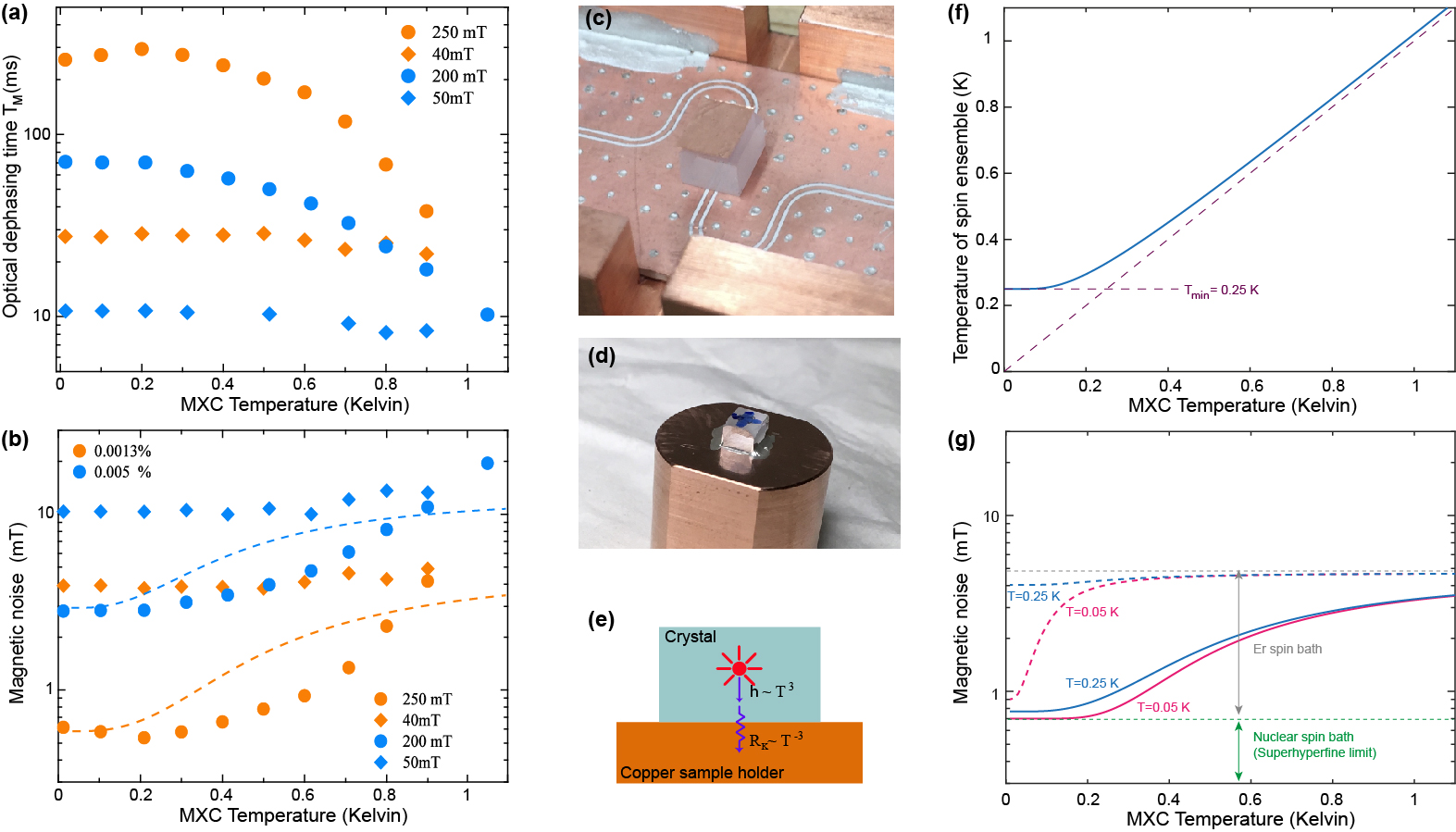}
\caption{(Color online) Optical phase memory time $T_M$ measured as the function of the measured temperature of the mixing chamber $T$ for 0.0013\% and 0.005\% Er:LYF crystals. \textbf{(b)} Magnetic noise $\delta B$ calculated from $T_M(T)$ data. Dashed lines represent simulated curve by using our model and parameters derived from $T_M(B)$ data. \textbf{(c)} The picture of the setup, where the crystal is thermally anchored through microwave substrate. \textbf{(d)} The picture of the setup where the crystal is thermally anchored via its gluing to the copper rod.\textbf{(e)} The picture illustrates the heat flow generated by laser radiation inside the crystal. The heat flow is impeded by the extremely low thermal conductivity below 1 Kelvin, which scales $\eta \propto T^3$, as well as thermal boundary resistance $R_K \propto T^{-3}$. \textbf{(f)} Solid curve illustrates behavior of $T_0$ as a function of $T$ if $T_{min}=0.25$~K. Dashed line corresponds to the case of $T_0=T$, or $T_{min}=0$~K. \textbf{(g)} Simulated magnetic noise as a function of $T_0(T)$ at minimal attainable temperatures of $T_{min}=0.05$ and 0.25~K. The noise parameters are used as for 0.0013\% sample. Dashed curves corresponds to 40~mT. Solid curves corresponds to 250~mT. The figure clearly displays the suppression of the magnetic noise at the weak field (40~mT) in the case of the deeper cooling to $T_{min}=0.05$~K.}
\label{Tm_versus_T}
\end{figure*}

The temperature dependence of optical dephasing time $T_M$ is shown in Fig.~\ref{Tm_versus_T}(a). The corresponding magnetic noise is calculated from the $T_M(T)$ data and plot in Fig.~\ref{Tm_versus_T}(b). For the simplicity we show the data measured at $0.05~$T and $0.2~$T for the 0.005\% sample and at $0.04~$T and $0.25~$T for the 0.0013\% sample. Above 1.5~K no echo is detected at pulse delay of $\tau=3-5~\mu$s for all fields. Knowing the minimal amplitude of the echo, which can be detected in our setup, we estimate $T_M\lesssim 1~\mu$s for the temperatures above 1.5 Kelvin. The temperature range $1-1.5~$K is hardly accessible for the dilution refrigeration because the circulation stability issues of $^3$He/$^4$He mixture. Therefore, the measurements are carried out only below 1~Kelvin. In this temperature range, $T_M$ measured at higher fields grows up with the temperature decrease and saturates around 0.2~K, so does the stretch factor $x$. For the weak fields $T_M$ stays nearly constant with a gentle decrease towards higher temperatures. In general, $T_M(T)$ and $\delta B(T)$ demonstrates similar behavior for both samples. 

Since the atomic ensembles stays at relatively high actual temperature of 0.2~K which exceeds the temperature of the mixing chamber $T$, at fields well below $100~$mT spin bath is unpolarized. In this case, the local magnetic noise as well as the optical dephasing time are mainly determined by the flip-flop dynamics of unpolarized electronic spins. At higher fields and at temperatures below 1~K, however, the electronic spin bath experiences substantial polarization leaving thus the local magnetic noise due to superhyperfine fields. 

In order to develop a deeper insight into the observed temperature dependence, we initially figure out the role of the thermal anchoring of the crystal. Below 1~Kelvin, thermal anchoring of isolators represent a major challenge even under moderate heat flow ($\sim\mu$W). In our experiment, the cw and pulsed laser illumination generates $\dot{Q}\sim0.1~\mu$W of heat inside the crystal. This heat should be transported through the bulk of the crystal and then be dissipated through the metal-isolator interface, see Fig.~\ref{Tm_versus_T}(e). Since the thermal conductivity is extremely low, i.e. $\eta(T)\propto T^3$ at ultra-low temperatures, a quite substantial thermal gradient appears inside the crystal. 

We estimate the temperature gradient to be $\nabla T = \dot{Q}/A\eta(T)$, where $A$ is the crystal area. If we assume for our crystal $\eta\sim10^{-7}~$W/cm$\cdot$K at 0.1~K, s.f.~\cite{Lounasmaa} then the temperature gradient generated by the fraction of $\mu$W is $\nabla T\sim 1~$K/cm. Therefore, it is not a big surprise that the temperature of the atomic ensemble is substantially higher than the temperature of the mixing chamber. In addition to that, thermal boundary resistance, which for the case of He-metal interfaces is the well known Kapitza resistance, impedes further the transport of the heat flow through crystal-metal interface. There, the temperature difference at the isolator-metal boundary is $\delta T=R_K \dot{Q}$, where $R_K\propto T^3$ is the thermal boundary resistance. However, it is hard to estimate such interface effect because cryogenic heat transport properties of laser crystals are unknown.

To explore which effect plays more important role in the heat transport through the crystal-sample holder, we run experiments with two types of thermal anchoring, as shown in Fig.~\ref{Tm_versus_T}(c),(d). In the first type, the crystal is placed on the microwave copper substrate (also discussed in Introduction). The copper sheet above the crystal is served for the better thermalization of the crystal via a copper screw. In the second type, we have used standard silver conductive paint RS Components. The similar thermal anchoring was made in our another optical experiments with SiV centers in diamonds below 1 Kelvin~\cite{Becher2017}. Both ways of the crystal thermalization yielded practically the same actual temperature of the crystal. Since the attained temperature was found to be independent on the type of boundary interface, we think that the thermal conductivity $\eta$ in the bulk of the crystal plays major role in the heat transport.

Due to the power dependence of the heat conductivity $\eta(T)\propto T^3$, the temperature $T_0$ of the spin ensemble largely deviates from the temperature of the mixing chamber only at very low temperatures. At higher temperatures, i.e. above 0.5~K, $T_0\approx T$. In principle, $T_0$ is a function of $T$.  We model it's behavior as $T_0(T)=T_{min}\coth\left({T_{min}/T}\right)$, where $T_{min}$ is the minimal temperature attainable for the spin ensemble. The behavior of $T_0(T)$ is shown in Fig.~\ref{Tm_versus_T}(f) for $T_{min}=0.25~$K. The dashed line in the same plot corresponds to the case of infinite thermal conductivity or $T_0(T)=T$. 

The influence of the actual temperatures $T_0$ of the spin ensemble on the magnitude of local magnetic noise is illustrated in Fig.~\ref{Tm_versus_T}(g). Here, the magnetic noise is simulated as a function of $T_0(T)$ accordingly to Eq.~\ref{dB_sum_TB_n}. The dashed lines correspond to the noise calculated for $B=0.05~$T for $T_{min}=0.05~$K and $0.25~$K, whereas the solid lines correspond to $B=0.25~$T and to the same temperatures. The noise parameters $\delta B^{(e})$ and $\delta B^{(n)}$ are taken for the 0.0013\% crystal. Due to the higher degree of polarization of electronic spins at higher fields the noise level attains the superhyperfine limit. At weak fields the superhyperfine limit can only be attained by the deep freezing of the ensemble to tens of millikelvins.

However, the proposed noise function only qualitatively describes the actual temperature dependence. Figure~\ref{Tm_versus_T}(b) displays a rather noticeable deviation of simulated curves from the data points for both samples. At lower temperature the deviation can be caused by the improper choice of the assumed function $T_0(T)$. However, at higher temperature the magnetic noise rapidly increases and exceeds the limit imposed by the unpolarized electronic spin bath. We think that such additional phase relaxation process may be produced by the thawing of the "frozen core". At higher temperature the reduced polarization of each erbium ion would result in a smaller "frozen core" radius. Nevertheless, we shall admit that a deeper insight into the nuclear spin dynamics in the presence of polarizing electronic spins can be reached only with the developing of specific theoretical models. Quite advanced theoretical models have already been developed for donors in silicon, NV-centers in diamond and quantum dots~\cite{Witzel2005, Witzel2012, Monteiro2015}. For the case of RE-doped crystals only simple model was numerically studied~\cite{Meltzer1992}. On the other hand, our model, which describes the influence of the magnetic noise around the clock transition, is oversimplified one, especially in the case of the spectral diffusion. Therefore, stimulated echo experiments are necessary in order to get a deeper insight into all magnetic and spin-lattice contributions~\cite{Thiel2010}. 

\section{Summary and conclusions}

In conclusion, we demonstrated cw and time resolved optical spectroscopy of isotopically purified $^{166}$Er:$^{7}$LYF$_4$ crystal below 1~Kelvin. Our fabricated crystals demonstrate  narrow optical FWHM linewidths down to 16 MHz. We found that at moderate magnetic field (below 0.3~Tesla), and ultra low temperatures (below 1 Kelvin) optical dephasing time attains $\sim10-100~\mu$s. By measuring the temperature and magnetic field dependence of the optical dephasing time we have studied the electronic and nuclear spin dynamics of Er:LYF system at sub-Kelvin temperatures. We found, that even for the very diluted sample with 0.0013\% doping concentration, the electronic spin noise plays the dominant role in the decoherence process at weak magnetic fields. The similar conclusion was derived in previous experiments with Er:YSO crystal at sub-Kelvin temperatures~\cite{Probst2015}. 

During the cw- and coherent spectroscopy, the crystal is heated by a laser radiation to the temperature higher than the temperature of the mixing chamber of the cryostat. For instance, at the base temperature of the fridge of 12~mK, the actual temperature of the spin ensemble is measured to be $T_0\simeq0.2~$K. Such thermal effect is associated with extremely low thermal conductivity of the crystal at ultra-low temperatures. Therefore, electronic spin fluctuations can be frozen out only by applying of moderate fields above $\sim 0.2~$T. In order to attain superhyperfine limit of local field fluctuations at weak ($\sim10~$mT) fields, the crystal has to be cooled down to $\sim10~$mK. Technically, such deep cooling of the sample may require fabrication of a crystal in the form of a single mode fiber~\cite{Tittel2017}. Owing to its small diameter, optical fibers can be better cooled to much lower temperatures than a large crystal~\cite{Hegarty1983,Staudt2006}. 
 
Overall, the presented experiment paves the way towards the implementation of long lived off-resonant Raman quantum memory by using narrowband solid state atomic ensembles at sub-Kelvin temperature range. Also, such atomic ensembles are very attractive for the realization of quantum converters between telecom C-band optical photons and superconducting quantum circuits~\cite{Longdell2015,Blum2015}. In this respect, we have demonstrated pulsed optical spectroscopy of erbium doped crystal at telecom C-band wavelength and close to the experimental conditions suitable for the operation of superconducting quantum circuits.

\section{Acknowledgment}

We thank M. N. Popova, A. V. Masalov for stimulating discussions, P.B greatly acknowledges discussions with Ch. Thiel, Th. B\"{o}ttger, W. Tittel, N. Sinclair, A. Tyryshkin and J. J. L. Morton. N.K. acknowledges support of G. Reicherz during NMR measurements. This work is supported by the Saarland University, Land of Saarland and DFG through the grant INST 256/415-1, BU 2510/2-1 and the RSF grant No. 14-12-00806.

\bibliography{coherence166}

\end{document}